\documentclass[bst/sn-mathphys-num]{sn-jnl}

\usepackage{graphicx}%
\usepackage{multirow}%
\usepackage{amsmath,amssymb,amsfonts}%
\usepackage{amsthm}%
\usepackage{mathrsfs}%
\usepackage[title]{appendix}%
\usepackage{xcolor}%
\usepackage{textcomp}%
\usepackage{manyfoot}%
\usepackage{booktabs}%
\usepackage{algorithm}%
\usepackage{algorithmicx}%
\usepackage{algpseudocode}%
\usepackage{listings}%
\usepackage{float}
\usepackage[super,sort&compress,comma]{natbib} 
\usepackage{array}
\usepackage{ulem}

\begin{document}

\title[Article Title]{\begin{center}Length-flexible strategies for efficient SERS\\ performance in gold-nanorod-gapped nanoantennas\end{center}}

\author[1]{\fnm{Sergio} \sur{F. Flores-Correa}}

\author[1]{\fnm{L. M.} \sur{León Hilario}}

\author[2,3]{\fnm{I. A.} \sur{Ramos-P\'erez}}

\author*[2,3,4]{\fnm{Andres} \sur{A. Reynoso}}\email{reynoso@cab.cnea.gov.ar}

\affil*[1]{\orgdiv{Facultad de Ciencias}, \orgname{Universidad Nacional de Ingenier\'ia}, \orgaddress{\street{Apartado 31-139, Av. Túpac Amaru 210}, \city{Lima}, \postcode{15304}, \state{Lima}, \country{Perú}}}

\affil[2]{\orgdiv{Instituto de Nanociencia y Nanotecnología (CNEA - CONICET)}, \orgname{Nodo Bariloche}, \orgaddress{\street{Av. Bustillo 9500}, \city{S. C. de Bariloche (RN)}, \postcode{8400}, \country{Argentina}}}

\affil[3]{\orgdiv{CNEA - CONICET}, \orgname{Centro At\'omico Bariloche and Instituto Balseiro}, \orgaddress{\street{San Carlos de Bariloche}, \city{R\'io Negro}, \postcode{8400}, \country{Argentina}}}

\affil[4]{\orgname{Universidad Nacional de Rio Negro}, \orgaddress{\street{San Carlos de Bariloche}, \city{Bariloche}, \postcode{8500}, \country{Argentina}}}

\abstract{Surface-enhanced Raman spectroscopy (SERS) using gold-nanorod-dimer nanoantennas has shown great potential in various applications. This reflects in their large values of the customary figure of merit of SERS: the enhancement factor (EF), which is essentially the fourth power of the electric field integrated at the gap, the location at which target molecules are to be sensed. However, fabrication errors in the nanorod lengths can lead to significant variations in the enhancement factor, resulting in performance limitations whenever low values of EF are encountered. Here, we report both design and procedural strategies to address this issue. First, we show that by reducing the nanorod diameter from 360 nm to 260 nm, the EF minima can be avoided for any nanorod length, mitigating the impact of fabrication errors. In addition, we explore the influence of incident wave polarization and orientation on the EF. Our simulations reveal that by tilting the excitation away from normal incidence, it is possible to substantially enhance EF under conditions that would otherwise exhibit low enhancement. In particular, this includes the case of 360 nm diameter. These findings expand the fabrication tolerance and broaden the range of usability of gold-nanorod-dimer nanoantennas, enabling more robust and reliable SERS performance. Importantly, we also show that these strategies also apply to nanoantennas with covered nanorod ends, which are of particular interest for realizing hybrid devices that combine SERS with electrical transport measurements.}


\maketitle

\section{Introduction}\label{Sec:Intro}

The miniaturization of electronic and photonic devices has driven the continuous exploration of light-matter interactions at the nanoscale~\cite{Weiner_book_2012, koya_2020,Weight_2023,Kamal_2022}. In this context, surface plasmons (SPs), collective electron oscillations that propagate along the interface between a metal and a dielectric, have emerged as a powerful tool for manipulating light and enhancing its interaction with matter~\cite{Schuller2010,Garcia_2011,Zhang_2012,dombi2020strong}. These properties make SPs particularly attractive for designing novel nanophotonic devices with applications ranging from sensing and imaging to the generation of localized heat sources in the treatment of cancer~\cite{Jain2007,dreaden2012golden,Zeng_2014,Duan_2021,kesharwani2023gold,Schuknecht2023}. In particular gold micro- and nano-structures can exhibit remarkable plasmonic properties at wavelengths ranging from the visible to the near-infrared region. For example the anisotropic geometry of gold nanorods can generate strong localized surface plasmon resonances (LSPRs). This geometrical control of the plasmonic behavior facilitates enhanced light confinement and scattering for advanced applications in sensing, imaging, and photonic devices.

An approach to harnessing SPs for light manipulation at the nanoscale involves gapped-metallic nanostructures due to their ability to efficiently concentrate and confine light within the gap region~\cite{Pavaskar_2012,Wei_2013,Zheng2021,Zhang_2022}. This localized light enhancement, or hot-spot, can be exploited for surface-enhanced Raman spectroscopy (SERS) \cite{SERSreview2025}, a technique that relies on the amplification of Raman scattering by molecules located on the nanostructure surface~\cite{Kneipp_1996,Lee2007,LERU200663}. In the field of SERS, the efficiency of nanoantennas is characterized by the well-known enhancement factor (EF). This figure of merit quantifies the amplification capability of the nanoantenna by integrating the fourth power of the electric field intensity over a defined surface within the hot-spot. The EF is a crucial parameter because it directly correlates with the degree of Raman signal enhancement achievable, thereby indicating the effectiveness of the nanoantenna in concentrating electromagnetic energy. By optimizing the design of nanoantennas to maximize the EF, one can significantly improve the sensitivity and performance of SERS for detecting phononic or other excitations of the target molecules~\cite{Kleinman2013,Amendola_2017,LeRu2024}. 


Here we focus on gold-nanorod dimers, namely, two gold cylinders, of diameters comparable to the working wavelength, separated by a gap of a few tens of nanometers. This platform presents three salient features: (i) its large gap area has the capacity to host higher amount of target molecules, (ii) the length of the constituent nanorods can reach the order of microns, allowing for the possibility of contacting both arms of the dimer enabling complimentary electrical transport measurements through the target molecules~\cite{ML19-Chen2009, Chen_Angewandte_2009, Cirera2022}, and  (iii) the multi-scale nature of the platform involves retardation effects and the availability of higher order modes~\cite{Payne2006,SchmidtPRB2008,Turner2010,YanQiu2022,Lidong}. We note that recent analytical efforts, devoted to the sub-wavelength version of this geometrical design, were obtained applying a quasi-static approximation~\cite{Downing2020}. Experimental interest includes cases in which either the transversal or all dimensions are much smaller than the wavelength~\cite{roa2024laser,Wu2014}. Here, on the other hand, since both the diameter and length of the nanorods are comparable to the wavelength, full wave computations are required~\cite{ML9-Dorfmuller2009,drai,dra,Goodman_1991,Ros2014,ML12-Zhang2019}. Thus, simulations become essential for the goal of identifying optimal geometrical parameters producing high EF values at the gap. Theoretical predictions resorting to discrete dipole approximation or finite elements methods, as the one we present in this work, have shown good agreement with experimental measured SERS in uncovered samples: see for instance Ref.~\cite{pedano,shuzho,AlexanderNanoLet2010}

In particular, previous theoretical investigations have found geometrical designs of covered dimers that do not degrade the SERS enhancement factor. These simulations---made for dimers with the far-from-the-gap edges covered by a layer of gold~\cite{IvanRSC2021, IVANPCCP2022} and for the experimentally explored case of diameter $360$ nm and normal incident $633$ nm-wavelength laser in uncovered dimers---identified optimal geometrical parameters reaching high EF values that even surpassed the largest ones obtained in the uncovered dimer. Thus, the platform has the potential to realize hybrid optic and electric molecule sensing. However, both for the covered and the uncovered cases, for some dimer lengths the EF can drop to very low values. This behavior presents a challenge as fabrication errors in the lengths of the dimer can potentially degrade the SERS signal.

In this work we address this issue by performing full wave simulations to compute the electric field enhancement factor in unexplored configurations. We report two strategies aimed at increasing the EF values in the presence of non-optimal dimer fabrication lengths. The first strategy involves working with samples with a reduced diameter value. We show that in this way the obtained EF profile becomes very good irrespective of the length of the dimer. Importantly, such benefit is already present for $260$ nm nanorods' diameter. This diameter value avoids the disadvantage of reducing too much the gap area and, at the same time, it can be fabricated resorting to the same methods successfully used to synthesize the $360$ nm-diameter-rod case of already proven samples~\cite{pedano, shuzho,ML18-Osberg2012,Osberg2011, Kim_2021}. Secondly, we also identify a methodological approach to avoid very low EF values even when the lengths of the sample would generate an EF minimum for the customary setup with diameter of $360$ nm. While the EF degrades by changing the incoming wave polarization away from the rod axis, we show that the enhancement factor can reach large values by slanting the $\mathbf{k}$ vector of the incoming wave towards such axis. Therefore, the presented two strategies allow overcoming previous limitations and enable SERS operation of gold nanorod dimers with greatly reduced geometrical constraints. Moreover, we show that the improvement also applies to covered designs thus paving the way to relaxing fabrication constraints of hybrid opto-electric molecular sensors. We remark that these successful strategies exploit the fact that here the geometrical dimensions are not negligible with respect to the wavelength. This becomes a resource as it opens the retardation regime in which the strategies (if carefully designed) allow for the excitation of distinct modes thus improving the enhancement factor without requiring tuning the nanoantenna length.

The paper is organized as follows: in Sec.~\ref{SC:Model} we introduce the definition and modeling of gold-nanorod-dimer nanoantenna configurations and present the strategies to address the problem. The enhancement factor results are presented in Sec.~\ref{SC:Results} for the cases mentioned above. Finally, in Sec.~\ref{SC:Conclusions} we conclude and discuss the main findings.

\begin{figure}[!t]
    \centering    \includegraphics[width=0.9\textwidth]{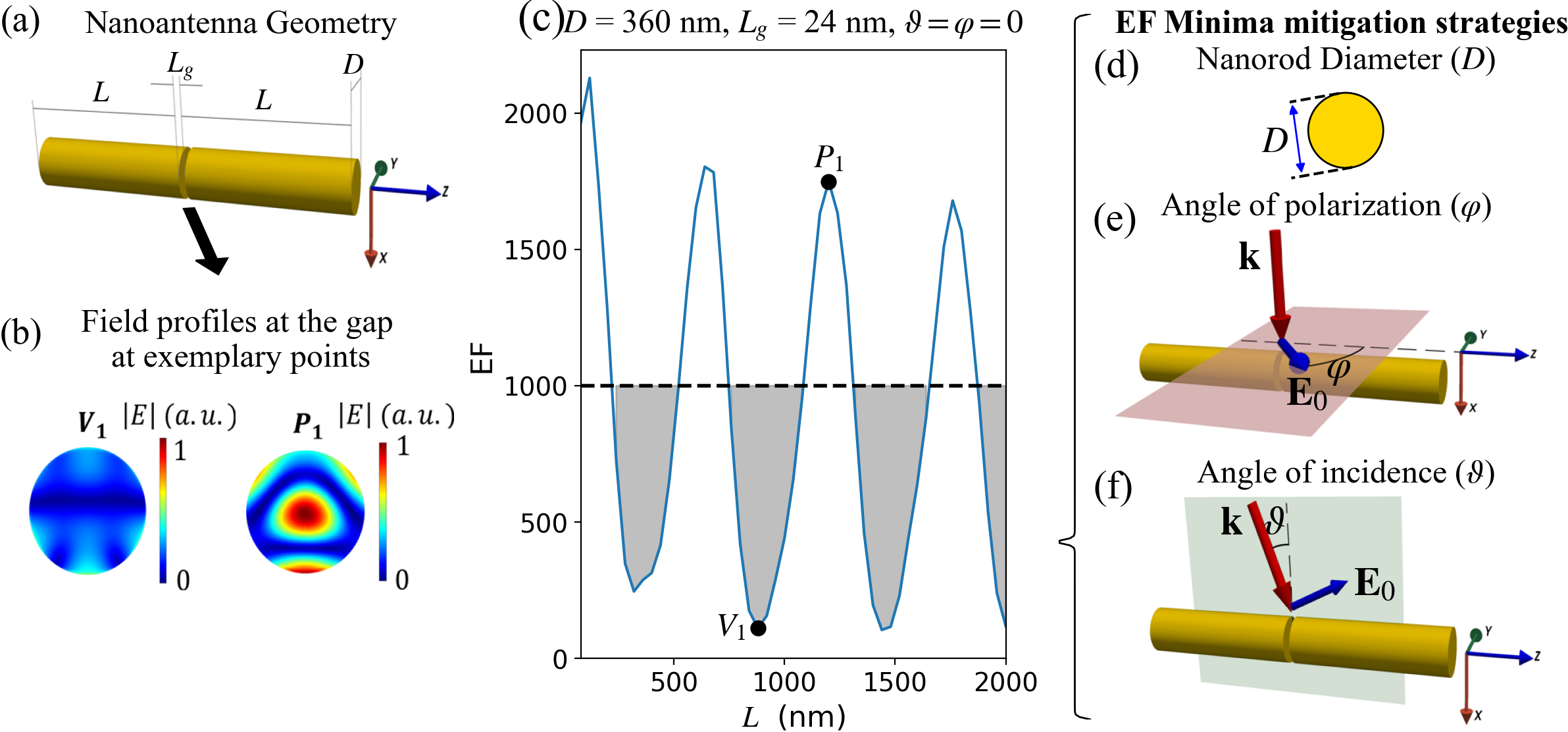}
    \caption{(a) Nano-antenna geometry: two $z$-axis-oriented-gold nanorods of diameter $D$ and length $L$ are separated by a gap length $L_g$ where target molecules are detected with SERS. Panels (b), (c) show simulations obtained for $D=360$ nm, $L_g=24$ nm and the customary case of a normal incident $\lambda_0=633$ nm $z$-polarized excitation from a Helium-Neon laser~\cite{pedano,shuzho}. (b) The SERS signal improves when the electric field profile at the gap becomes very large (see $P_1$ in (c)), however, small geometrical differences may lead to low field profiles (see $V_1$ in (c)). (c) This produces an oscillating dependence as a function of $L$ of the enhancement factor, $\text{EF}\propto |\mathbf{E}|^4$, a figure of merit of SERS spectroscopy. Samples fabricated with $L$ values at the shaded region would develop degraded values of $\text{EF}$. Studied strategies for obtaining large $\text{EF}$ at all lengths $L$, without the need of changing the He-Ne $\lambda_0=633$ nm source, are: (d) Sample diameter modification; (e) tuning the angle of polarization $(\varphi)$ away from the $z$-axis while keeping the wavevector $\mathbf{k}$ at normal incidence; or (f) introducing an inclination  angle $(\vartheta)$ to the incoming wave while keeping both the electric field and the wavevector in the $z-x$ plane. }
    \label{fig:setup}
\end{figure}

\section{Definition and modeling of gold-nanorod dimer experimental configurations}
\label{SC:Model}

\subsection{Geometry and optical setup}
We focus our theoretical study on realistic platforms considering as a starting point the uncovered gold nanorod-dimer devices already realized experimentally in previous studies~\cite{pedano,shuzho}. The required fabrication techniques, described with detail in previous works are widely available~\cite{ML15-Braunschweig2010, ML14-Hurst2006,ML17-Lim2016,ML19-Chen2009}. They involve on-wire lithography (OWL) with selective chemical etching of sacrificial segments~\cite{Banholzer2009},  allowing for the fabrication of the basic design depicted in Fig.~\ref{fig:setup}(a). The dimer consists of two gold nanorods, or nanocylinders of diameter $D$, characteristic length $L$, and a gap length $L_g$ separating them. As mentioned above, the preferred values of $L$ range around $1\mu$m allowing the contacting of the device by depositing gold at the edges of the rods, i.e., far from the gap area. For the same reason the diameter $D$ can not be too small, and was taken $360$ nm in the previous studies. This also ensures a sufficiently large gap area to accommodate a large amount of target molecules. Hot-spots of the electric field at the gap that enhance the sensibility of Raman spectroscopy are obtained for gap lengths $L_g<30$ nm. In particular, following previous experimental successful realizations, we work with $L_g=24$ nm which is apt to host DNA molecules that can touch both cylinders enabling eventual electric transport measurements that would complement the SERS measurements.

Simulations of the geometric and excitation parameters are key to optimize the ability of the device for Raman spectroscopy. Previous studies focused on the normal incidence excitation (the wavevector $\mathbf{k}$ of the incoming wave is parallel to the $x$ direction in Fig.~\ref{fig:setup}(a)) with $\mathbf{E}$-field polarization along the nanorods cylinders (the $z$ direction). In particular, the assumed incoming drive is a continuous plane wave from a Helium-Neon laser operating at a wavelength of $\lambda_0=633$ nm, a source widely available in spectroscopy laboratories. This specific wavelength, which is known to suit well for gold-based plasmonic devices, has been chosen in prior experimental and theoretical research~\cite{pedano,shuzho,IvanRSC2021, IVANPCCP2022}, enabling direct comparison of the simulated results. To maintain these advantages in this study we keep this wavelength fixed and proceed to explore other conditions, as detailed in the next section.

\subsection{Obtaining the SERS Enhancement Factor}
Considering the weak coupling approximation~\cite{YAMAMOTO201481}, the SERS effect arises from a two-step enhancement: a local field enhancement $|M_1|^2$ during the excitation process and radiation field enhancement $|M_2|^2$ during re-emission~\cite{LERU200663}; hence, the total SERS EF for a single molecule (SMEF) can be expressed as:
\begin{equation}
    \text{SMEF} = |M_1(\omega_L)|^2|M_2(\omega_R)|^2,
\end{equation}
with $\omega_L$ and $\omega_R$ the excitation and radiation frequencies---where $\omega_R=\omega_L\pm\omega_{ph}$ with the minus (plus) sign for a Stokes (an anti-Stokes) Raman process involving a phonon excitation in the target molecule of frequency $\omega_{ph}$. Leveraging the optical reciprocity theorem and assuming a single molecule SERS condition both factors can be expressed in terms of the local (or near) field $\mathbf{E}_{\text{loc}}$ and the incident field $\mathbf{E}_0$ as 
\begin{equation}
    \text{SMEF} = \bigg|\frac{\mathbf{E}_{\text{loc}}(\omega_L)}{\mathbf{E}_0(\omega_L)}\bigg|^2\bigg|\frac{\mathbf{E}_{\text{loc}}(\omega_R)}{\mathbf{E}_0(\omega_R)}\bigg|^2.
\end{equation}

For a sufficiently high incident frequency, the difference in $|\mathbf{E}_{\text{loc}}|$ at the excitation and re-emission frequencies becomes negligible due to the relatively small energy of phonons. Therefore, one may consider $|\mathbf{E}_{\text{loc}}(\omega_L)|  \approx |\mathbf{E}_{\text{loc}}(\omega_R)|$, which leads to the well-established $|\mathbf{E}_{\text{loc}}|^4$ approximation.

In this work, we calculate the EF to characterize the efficiency of the SERS effect~\cite{LERU200663, librochapter5}. We employ the $|\mathbf{E}_{\text{loc}}|^4$ approximation and average it over a transversal section inside the gap:
\begin{equation}
\text{EF} = \frac{\int_{S}|\mathbf{E}_{\text{loc}}|^4/|\mathbf{E}_0|^4dS}{\int_S dS},
    \label{eq:EF}
\end{equation}
with $S$ a circular surface inside the gap region located $2$ nm from one of the faces of a cylinder, see for example the computed electric field profile in Fig.~\ref{fig:setup}(b).

To compute the electric field of the full system and proceed to obtain the EF we perform numerical simulations using a standard package of finite element method, Comsol multiphysics, to solve Maxwell's equations that accurately model the electromagnetic behavior of the nanorod dimers. A 3D model replicating the geometry in Fig.~\ref{fig:setup}(a) setup was carefully constructed with a mesh that considered skin depth effects. This was achieved by introducing a separate layer of $30$ nm on both the inner and outer volume near the surface of the nanorod with a minimum mesh size of $3$ nm along the normal direction and at least $20$ nm along the longitudinal direction. In the gap region, a fine mesh with elements in arithmetic progression ensured proper resolution of the electric field within the gap, down to a $1$ nm resolution near the cylinder's surfaces. Several test simulations were performed with different numbers of elements within the gap to determine the optimal configuration considering the trade off in accuracy and computational cost. The optimal value was determined to be $20$ elements along the cylinder's axis. To truncate the infinite domain, a perfectly matched layer (PML) boundary condition was applied surrounding the geometry, positioned two wavelengths away from it, which would serve to absorb all incident light and avoid back reflections to the system. Simulations were conducted with the relative dielectric constants of gold taken from the work by Johnson and Christy~\cite{PhysRevB.6.4370} where, at an incident wavelength of $\lambda_0=633$ nm, the real and imaginary parts are $\epsilon_1=-11.753$ and $\epsilon_2=1.2596$, respectively. 

Before proceeding to explore novel configurations we verified that obtained results were in agreement with previous studies for nanorods of $360$ nm performed with the same finite element method~\cite{IVANPCCP2022} and also with results obtained by DDSCAT 7.3 implementation of the Discrete Dipole Approximation (DDA) method~\cite{dra,pedano,IvanRSC2021}. This comparison includes and is not limited to the amplitude of the electric field, value of the enhancement factor and charge density profile, which yielded excellent agreement, validating the accuracy of our simulations. It is worth noting that dealing with the $\vartheta\neq 0$ or the $\varphi\neq 0$ cases (see Fig.~\ref{fig:setup}(e)-(f) and the definitions in the next section) involves the breaking of spacial-symmetries in the solutions demanding a sparse matrix occupying up to $230$GB of memory. For longest dimer nano-antenna the mesh contained up to $10^7$ complex-valued degrees of freedom and the typical simulation time of the full electromagnetic field for a single parameter could take up to $12$ hours.

\subsection{Strategies to overcome low values of EF regardless of $L$}
\label{SC:sub:strategies}
Figure~\ref{fig:setup}(c) presents the calculated enhancement factor as a function of cylinders' length $L$ for the above discussed case of normal incidence and $z$-polarized field (for fixed $D=360$ nm and $L_g=24$ nm). The peaks of EF arise whenever the excitation drives a plasmonic mode of the dimer that hosts a hotspot at the gap. Naturally, this is a size dependent effect since the dimensions of the coupled nanorods are comparable to the fixed laser wavelength $\lambda_0$. For thin (compared to the wavelength) gold strips an effective Fabry-Perot picture has been successful in explaining the surface plasmon polaritons (SPPs) resonances in terms of $L$~\cite{Barnard2008FPthin,ML9-Dorfmuller2009}. For thicker nanorods one can attempt to apply this simplified picture~\cite{shuzho,Gordon:09} to the single segment case. In this way longitudinal modes of the system are associated to Fabry-Perot-like resonances of the SPPs, with wavelength $\lambda_\text{SPP}\approx600$ nm~\cite{pedano}, that travel back and forth the nanorod and interfere constructively. For a Fabry-Perot interferometer, the resonances arise for lengths differing in multiples of $\frac{ \lambda_\text{SPP}} {2}$. The normal incident $z$-polarized driving only excites odd parity of the modes~\cite{shuzho}, and thus the signatures in the signal appear at values of $L$ distanced by multiples of the full $\lambda_\text{SPP}$. In fact, even the $D=360$ nm case of two nanorods coupled by the gap in Fig.~\ref{fig:setup}(c) develops EF peaks separated by $\lambda_\text{SPP}$. However, as we show in Sec.~\ref{SC:Results} the SPP modes of the full dimer in general can be very different to those expected with simple Fabry-Perot interferometer arguments. This is due to the losses of the metals, virtually absent in photonic Fabry-Perot interferometers and, most importantly, to the fact that retardation effects along the non-negligible perimeter (the transverse direction to the $z$ axis) are also present.

The shaded region in Fig.~\ref{fig:setup}(c) illustrates the main issue we address in this paper, namely, that for significant ranges of the $L$ parameter the EF value can be greatly degraded. Indeed, Fig.~\ref{fig:setup}(b) shows how the norm of the electric field at the gap drastically drops with a small change of $L$ from a peak to a valley of EF. Note that in practice $L$ could be off from an optimum value either due to fabrication errors or due to lack of information of the EF profile. Moreover, when aiming at complementary electric transport measurements, the edges of the rods are covered by an additional layer of gold, and the resulting length can also be affected. This motivates us to unveil ways to recover high enhancement in the SERS signal without the need to fine-tune $L$. We note that, given the system size to wavelength ratio in these devices, the already mentioned retardation effects can become a \emph{resource} for exciting other modes that could provide enhancement of the near field at the gap. Such effect can be exploded either by direct geometry modifications of the initial setup or by modifying the excitation configuration~\cite{Pelton2013book}. In Sec.~\ref{SC:Results} we present results in this direction based on exploring the following strategies to overcome the problem of the EF minima:

\paragraph*{Diameter change.} In Fig.~\ref{fig:setup}(d) one envisions working with nanorods of diameters smaller than $360$ nm as long as: (i) the eventual covering contact can still be realized, and (ii) the net gap area does not decrease too much for it to accommodate significant amount of target molecules. For example, $D=260$ nm is a reasonable value for this purpose. In fact, anodic aluminum
oxide (AAO) membranes templates for growing nanorods with $260$ nm diameter are readily available~\cite{privcomMLPedano}.

\paragraph*{Polarization angle.} The incoming electric field polarization is contained in the $yz$-plane (colored blue in Fig.~\ref{fig:setup}(e)) as the incident wavevector $\mathbf{k}=\frac{2\pi}{\lambda_0}\hat{x}$ is kept normal. The angle of polarization $\varphi$ is defined as the angle between the electric field vector of the incoming wave $\mathbf{E}_0=|\mathbf{E_0}|\hat{z}\cos{\varphi}-|\mathbf{E}_0|\hat{y}\sin{\varphi}$ and the $z$-axis. Note that former studies with $z$-polarized case correspond to the $\varphi=0$ case.

\paragraph*{Incidence angle.} Here the $zx$-plane (colored green in Fig.~\ref{fig:setup}(f)) contains both the $\mathbf{E}_0$ and $\mathbf{k}$ vectors as they remain orthogonal. The angle of incidence is defined as the angle between $\mathbf{k}$ and the $x$-axis. Explicitly, in this configuration the incoming wavevector and electric field polarization vectors read
\begin{equation}
\mathbf{k}=\frac{2\pi}{\lambda_0}(\hat{x}\cos{\vartheta}+\hat{z}\sin{\vartheta})
\,,\,\,\,\,\,\,
\mathbf{E}_0=|\mathbf{E}_0|(\hat{z}\cos{\vartheta}-\hat{x}\sin{\vartheta}).
\label{eq:inclination}
\end{equation}
Then this setup coincides with former studies of $z$-polarized electric field and normal incidence for $\vartheta=0$. 

This setup naturally assumes that the angle in which the laser hits the sample has to be modified. One can envision, however, realizing this situation by leveraging recent advances in dynamic beam steering systems, which enable continuous adjustment of the emitted radiation angle without physical realignment of the incident laser \cite{Dutta,Busschaert,Bhaskar}. We do not discuss further this possibility that could provide a pathway to implement the oblique incidence configuration in more compact setups.

\begin{figure}[!t]
    \centering
    \includegraphics[width=0.9\textwidth]{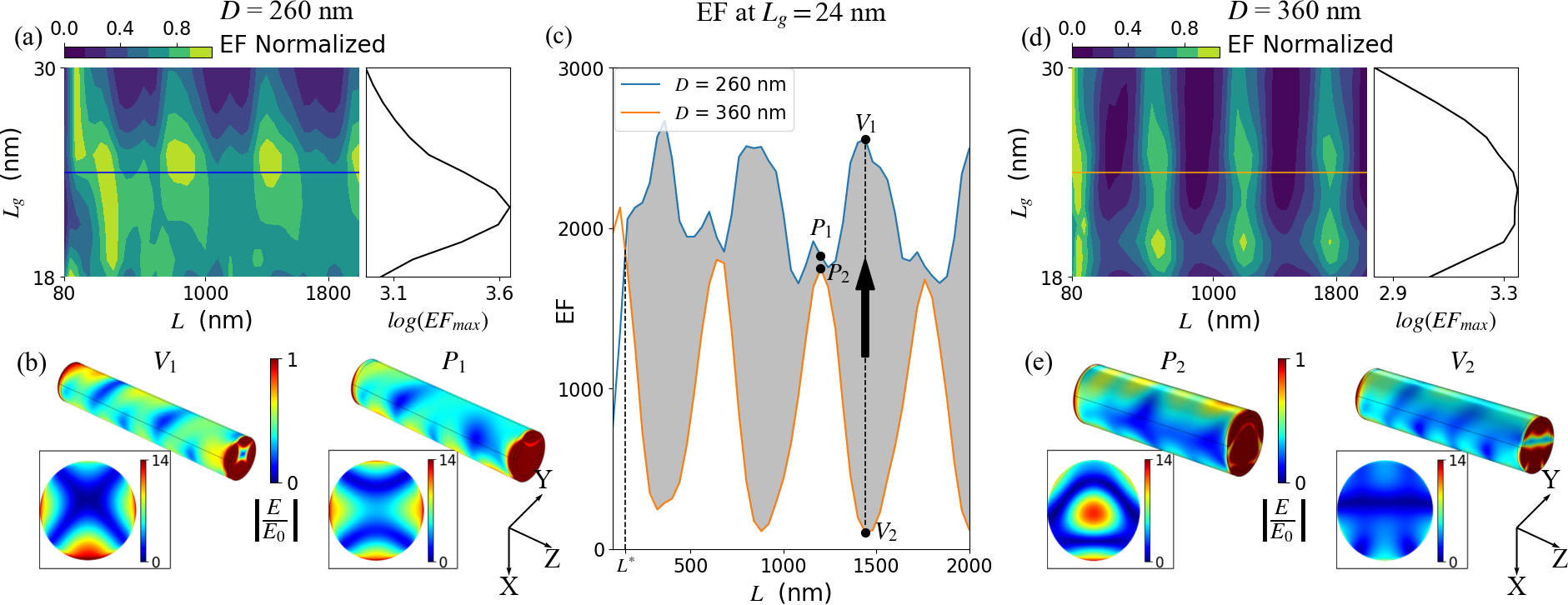}
    \caption{SERS Enhancement Factor versus $L$ and $L_g$ for (a) $D=260$ nm and (d) $D=360$ nm. The colormaps are normalized at each $L_g$ with respect to the maximum EF, $EF_{max}$, shown (in log 10 scale) at each side left panel.(c) EF versus $L$ cuts at $L_g=24$ nm; the arrow and the shaded area highlight the EF improvement of the $D=260$ nm case at all lengths $L$ at which $D=360$ nm develops low EF values. Panel (b) [(e)] shows, for the $D= 260$ [$D=360$ nm] case, the electric-field norm maps at the gap and on a cylinder (the other cylinder is symmetric) at the exemplary positions $V_i$ ($L=1440$ nm) and $P_i$ ($L=1200$ nm) indicated in (c).} 
    \label{fig:Results1}
\end{figure}

\section{Results}\label{SC:Results}

\subsection{Avoiding Enhancement Factor minima using samples with smaller nanorod diameters}
\label{sc:diameter}
Following the first strategy listed in Sec.~\ref{SC:sub:strategies} we keep normal incidence and $z$-polarized incoming field and proceed to change the diameter away from the customary value of $360$ nm. The EF profiles for a sweep of diameter is included in the Appendix. We select the $D=260$ nm case as an exemplary diameter that produces large EF values for all relevant $L$ as shown in Fig.~\ref{fig:Results1}(a) within the gap length range around the target working condition of $L_g=24$ nm. The non-trivial dependence with $L_g$ manifests that retardation effects around the perimeter dimension can produce complex multimode effects. In particular, the EF profiles versus $L$ for this diameter indicate the excitation of more modes around $L_g=24$ nm leading to large EF at all $L$. On the other hand the $D=360$ nm case in Fig.~\ref{fig:Results1}(d) mostly shows a unique family of modes and thus $L$ need to be tuned to avoid the minima of the EF in agreement with Fig.~\ref{fig:setup}(c). Figure~\ref{fig:Results1}(c) highlights the EF versus $L$ for the cuts at the working gap length $L_g=24$ nm: the computed EF for $D=260$ nm surpasses the EF for $D=360$ nm for all $L$ values.   

Figures~\ref{fig:Results1}(b) and (e) show the patterns of electric field at fixed values of $L$ at exemplary conditions at which the $D=360$ nm case had a peak (a valley) EF value at points labelled $P_i$ ($V_i$) in Fig.~\ref{fig:Results1}(c). The maps show the norm of the field on the surface of a single nanorod---the other one is symmetric given that here $\vartheta=\varphi=0$. As the wavelength to diameter ratio changes the arrangement of the plasmon modes is affected, something that produces distinct shapes of field at the gap included in Fig.~\ref{fig:Results1}(b) and (e)---see Appendix~\ref{SC:Appendix1}~\cite{Lidong}. This observation is consistent with the obtention of different EF profiles as a function of $L$ when operating at the regime in which $D$ is not negligible with respect to $\lambda_0$.

\begin{figure}[!t]
    \centering
\includegraphics[width=0.9\textwidth]{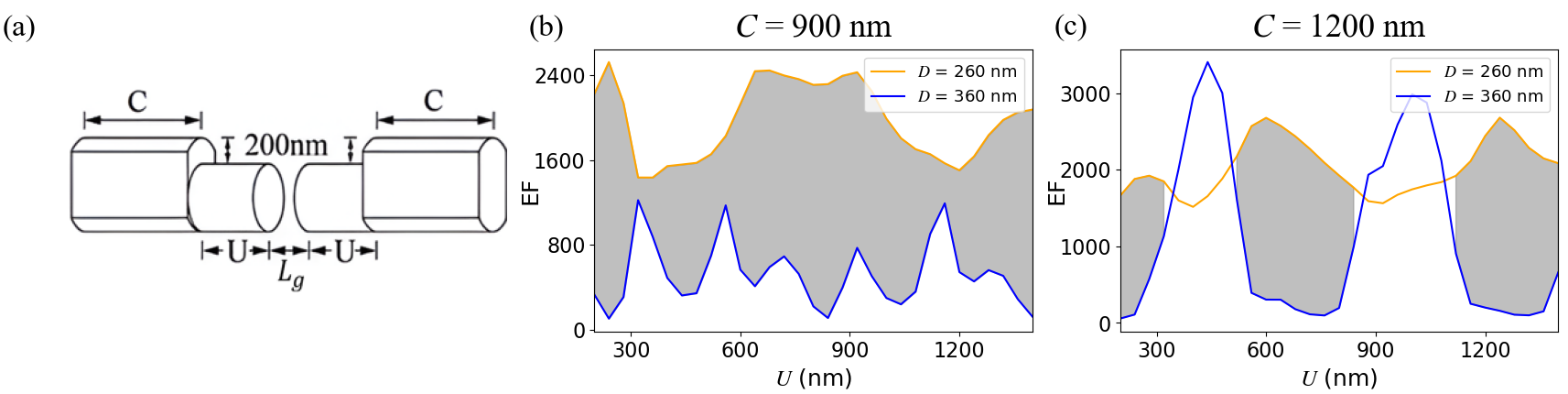}
\caption{(a) Sketch of the covered case studied in previous studies. Enhancement Factor at the gap versus $U$ for covered length (b) $C=900$ nm or (c) $C=1200$ nm---for the customary excitation of normal incidence and $z$-polarized field as in Fig.~\ref{fig:Results1}. The sample with $D=260$ nm presented here achieves large EF values regardless of the value of the uncovered section $U$ for both covered lengths. The shaded areas highlight the ranges of $U$ values at which a sample with $D=360$ nm would obtain poorer SERS responses.}
    \label{fig:Results2}
\end{figure}

We proceed to test whether or not the success of this strategy to mitigate EF minima also apply to the case in which the outer edges of the nanorods are covered. As already mentioned, these covered geometries are motivated by the prospect of electrically contacting the dimer. Here, though the covered designs are not the focus of this work, we take the sample geometry studied in previous works~\cite{IvanRSC2021,IVANPCCP2022} shown in Fig.~\ref{fig:Results2}(a) and compute the EF for a few exemplary cases with different covered lengths $C$. Note the uncovered length $U$ of both nanorods has diameter $D$ and are the regions in contact with the gap, therefore, the expectation is that by setting $D=260$ nm one could avoid the appearance of EF minima even in a covered design. Figure~\ref{fig:Results2}(b) presents the EF profiles versus $U$ for $C=900$ nm demonstrating that, indeed, the $D=260$ nm dimer obtains large EF without requiring fine tuning $U$. Note that at such value of $C$ the already studied $D=360$ nm case presents moderate to low EF values at all $U$. In contrast, for $C=1200$ nm the $D=360$ nm case presents large values of EF for some particular $U$ but at the same time the EF drops at ranges of $U$, i.e., the behavior we aim to avoid. This case is shown in Fig.~\ref{fig:Results2}(c) and illustrates that the $D=260$ nm sample develop, as desired, large EF values for all values of $U$. Note this is true even if the $D=360$ nm dimer could outperform the $D=260$ nm one at some particular values of $U$. Importantly, the success shown in Fig.~\ref{fig:Results2} for two characteristic value $C$ qualitatively holds in the general case.           

In summary, the requirement to tune the length $L$ (or $U$) is no longer required to obtain large EF values, and this strategy is useful both for the uncovered case and for the covered cases. It has the advantage that it maintains the normal incidence and $z$-polarized excitation, but it involves working from the start with a diameter different than the original $360$ nm case. In the following sections we proceed to explore changes in excitation configuration to explore the possibility of obtaining good SERS performance without having to tune $L$ (or $U$) even for the $360$ nm diameter case.

\subsection{Enhancement Factor behavior with polarization}\label{sc:pola}

\begin{figure}[!t]
    \centering    \includegraphics[width=0.9\textwidth]{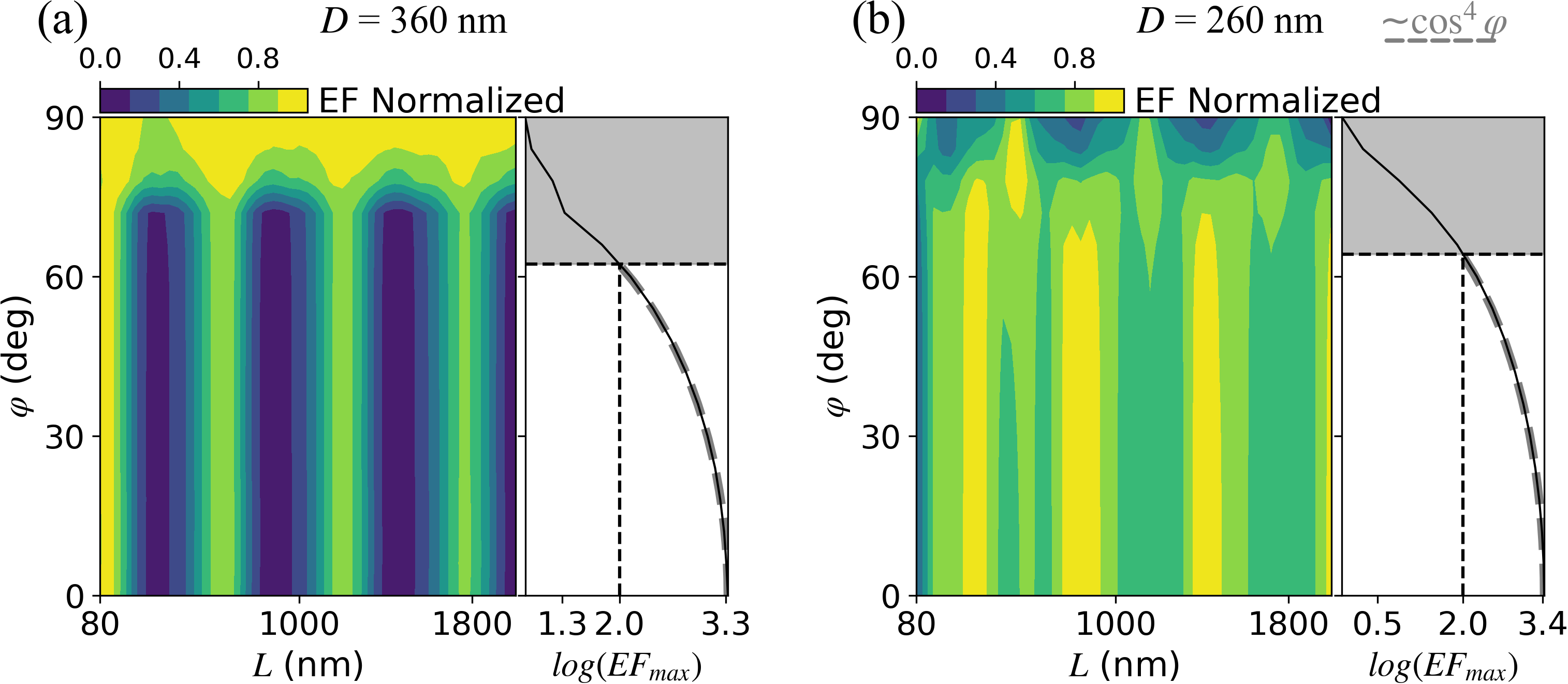}
    \caption{Simulations for (a) $D = 360$ nm and (b) $D = 260$ nm of the enhancement factor versus $L$ for normal incidence excitation with variable polarization angle $\varphi$ as sketched in Fig.~\ref{fig:setup}(e). For each $\varphi$ the profiles versus $L$ are normalized to the maximum value shown (in log 10 scale) at the left of each colormap. The dashed lines are $EF_{max}(\varphi=0)\cos^4\varphi$ (in log 10 scale) which are found to fit very well the numerical obtained EF maxima profile.}
    \label{fig:Results3}
\end{figure}

Following the plan described in Sec.~\ref{SC:sub:strategies} we now focus on strategies that can be applied at the measurement stage. We first investigate the EF behavior versus $L$ when the normal incident wave changes its polarization an angle $\varphi$ away from the axis of the nanorods, as sketched in Fig.~\ref{fig:setup}(e). We find that switching the polarization away from $z$-axis is detrimental for SERS spectroscopy purposes. This is summarized in Fig.~\ref{fig:Results3} where for two exemplary diameters: (i) the patterns maintain the structure of the $z$-polarized case ($\varphi=0$) and (ii) the maxima of EF (the $EF_{max}$ at the sidepanels in log 10 scale) decay as $\varphi$ grows. Since EF values below 100, corresponding to $\theta>72^{\circ}$, are not relevant for our purposes, they are represented in gray in these sidepanels.

The side-panels of Fig.~\ref{fig:Results3}(a) and (b) include dashed lines with $\cos^4\varphi$-proportional profiles (in log 10 scale) that successfully fit the numerically computed profiles of the EF maxima versus $\varphi$. This agreement is consistent with the electric field norm at the gap being modulated as $|\mathbf{E}(\varphi)|\approx|\mathbf{E}(\varphi=0)|\cos\varphi$. This means that only the component of the incoming field aligned with the $z$-axis is able to excite the surface plasmons that generate a large EF at the gap. Thus, maintaining a large component of polarization along $z$ is the most effective strategy for SERS purposes~\cite{roa2024laser}. This result is in fact expected from symmetry arguments, as a static field along $z$ would maximize the surface charge of opposite signs that would accumulate on the two circular sections adjacent to the gap region. Clearly, a reminiscence of such capacitor-like effect that maximizes the field at the gap is maintained in the case of a dynamic $\mathbf{k}\perp \hat{z}$ excitation with the incoming field $\mathbf{E}_0$ polarized along $z$. In sub-wavelength dimer samples, the dipoles of the two particles couple and hotspots at the gap develop more efficiently for this polarization. This polarization also maximizes the SERS efficiency when the system is not in the quasi-static regime.

\subsection{EF dependence with the angle of inclination}\label{sc:strategy3}
\begin{figure}[!t]
    \centering
\includegraphics[width=0.9\textwidth]{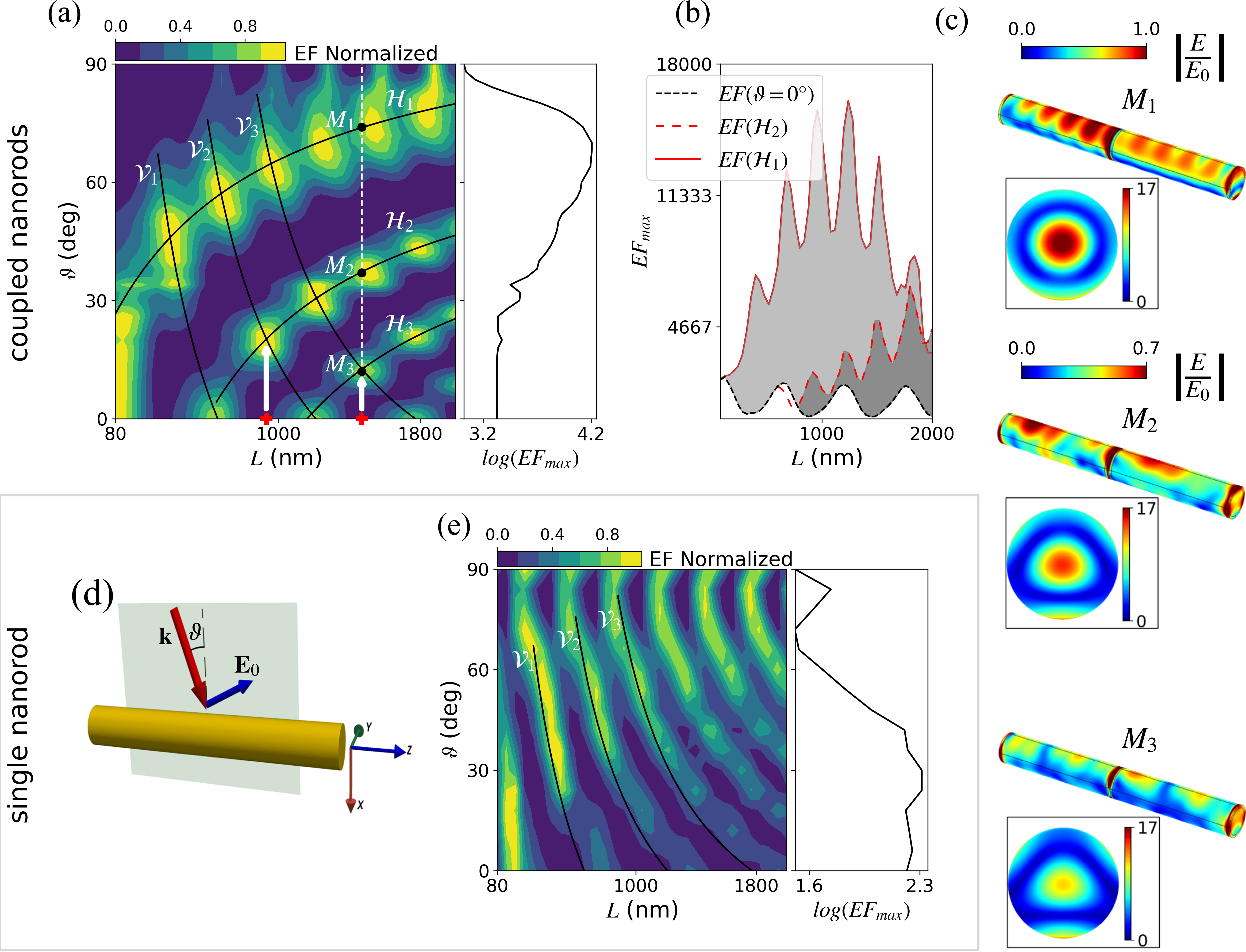}
    \caption{Mitigation of EF minima by adding $\vartheta$ at the measurement stage following the setup in Fig.~\ref{fig:setup}(f). (a) EF as a function of the nanorods' length $L$ and inclination angle $\vartheta$ for the $D=360$ nm case with $L_g=24$ nm. For each $\vartheta$, the profiles versus $L$ are normalized to the maximum values, $EF_{max}$, shown (in log 10 scale) at the colormaps' side-panels. The white arrows highlight that by increasing $\vartheta$ the EF be enlarged, this is shown at exemplary $L$ cases (red crosses) at which SERS spectroscopy is inefficient at the $\vartheta=0$ case. (b) EF at the gap versus $L$ for $\vartheta = 0$ and along the two lines, $\mathcal{H}_1$ and $\mathcal{H}_2$. Shaded areas highlight the ranges of $L$ where the EF obtained with nonzero $\vartheta$ along the $\mathcal{H}_i$ surpasses those of the normal incidence excitation ($\vartheta=0$) of Fig.~\ref{fig:setup}(c). (c) Surface distribution of the normalized electric field $|E/E_0|$ for the three conditions $M_1$, $M_2$, and $M_3$ shown in (a) for a nanorod dimer with diameter $D=360$ nm. Insets provide cross-sectional views of the field profiles inside the gap. (e) Enhancement factor as a function of $L$ and $\vartheta$ for the case of a single nanorod sketched in (d); the $\mathcal{V}_i$ lines from the dimer case of (a) are included.}
\label{fig:Results4}
\end{figure}

We proceed to explore the third strategy presented in Sec.~\ref{SC:sub:strategies}, i.e., changing the excitation direction away from the $\mathbf{k}\perp \hat{z}$ condition, as sketched in Fig.~\ref{fig:setup}(f). The angle of inclination or tilting, $\vartheta$, is increased while ensuring that the polarization of the electric field along the $z$ axis is maximum. The reason to do this is, as seen in the former section, that the polarization along $z$-axis is the one that most naturally drives large values of the field at the gap. This condition is achieved with the $\mathbf{k}$ and $\mathbf{E}_0$ orientations given in Eq.\eqref{eq:inclination}, where as we take $\mathbf{k}\cdot\hat{y}=0$ then $\mathbf{E}_0$ is also contained in $z$-$x$ plane for maximizing the projection along $z$\footnote{Simulations demonstrate (not shown) that the $y$-quadrature is inefficient to excite modes with large EF at the gap. This includes the case of circular polarization, in which only the component along the axis considered in the text (the $\mathbf{E}_0$ defined in Eq.\eqref{eq:inclination}) contributes to excite modes with enhanced near-field amplitudes at the gap.}.

Note that for $\vartheta\neq 0$ the left to right cylinder symmetry of the electric field is broken and therefore the EF, as defined in Eq.\eqref{eq:EF}, presents differences if computed over a surface $S$ near the left or the right nanorod. In the Appendix we show that both patterns of EF versus $L$ are qualitatively the same and the quantitative differences are small. In this section, to account for this left/right asymmetry the EF presented at every point (when $L$ and $\vartheta$ change) is the minimum of the two values, thus providing a conservative estimate of the gain for different orientations.

As the $D=260$ nm case develops good EF for all $L$ case here we do not consider such case. We focus on the $D=360$ nm SERS enhancement factor response as a function of $L$ and $\vartheta$ for the uncovered dimer of Fig.~\ref{fig:setup}(a) with $L_g=24$ nm. The colormap in Fig.~\ref{fig:Results4}(a) shows that as $\vartheta$ is varied the EF pattern gets modified. This is a good indication since the structural alteration of the EF profile versus $L$, compared to the normal incidence case, is a prerequisite to obtain large EF at ranges of $L$ which otherwise produce low enhancement for $D=360$ nm, see shaded regions in Fig.~\ref{fig:setup}(c). In Fig.~\ref{fig:Results4}(a), we observe that configurations yielding low enhancement in the normal incidence case, marked with red crosses, can achieve optimal EF values by adjusting the angle of incidence. These transitions from EF local minima to  EF local maxima are highlighted with white arrows. Remarkably, the strategy is successful since whenever one encounters a $L$ value that produces low EF at normal incidence one can proceed to increment the $\vartheta$ angle thus enhancing the EF even to values that surpass those expected if $L$ were optimized to the normal incidence case.

In this way, the suggested procedure involves profiting from the presence of the local maxima in the EF versus $(L, \vartheta)$ map in Fig.~\ref{fig:Results4}(a). In order to facilitate the analysis it is useful grouping the EF maxima in two ways. First we group them in \emph{Vertical Families}, denoted as $\mathcal{V}_i$. Along these lines $|\frac{d\vartheta}{dL}|$ is large as the length is kept as fixed as possible. In Fig.~\ref{fig:Results4}(e) we include the $\mathcal{V}_i$ lines of the dimer case over the map of the (much smaller) EF computed over a circular surface at the end of the single nanorod case depicted in Fig.~\ref{fig:Results4}(d). This helps to visualize that the EF maxima of the single nanorod case arrange in a similar fashion that the $\mathcal{V}_i$ lines. Note that in the dimer case the EF at the gap switches drastically on and off along these lines, whereas in the single nanorod case the EF does not present strong oscillations. This fact reflects an interference pattern arising from the SPPs coupling between the two nanorods. A second natural way to group the EF maxima is in what we call \emph{Horizontal Families}, labeled as $\mathcal{H}_i$. Along these lines the $\frac{d\vartheta}{dL}$ is moderate to low and the EF, in between neighboring maxima conditions, does not drop too much. This means that the valleys along these lines also produce good SERS enhancement factor. We note that these families are absent in the single nanorod case, evidencing that the coupling of the two nanorods and the resulting interference effects are responsible for the strong enhancement developed in the dimer case. The specific dependence with $L$ and $\vartheta$ of both the $\mathcal{H}_i$ and $\mathcal{V}_i$ families are provided in the Appendix.

To verify that this strategy can enlarge the EF at all values of $L$ it is useful to study the enhancement profile along the different $\mathcal{H}_i$ lines. This is presented in Fig.~\ref{fig:Results4}(b), where shaded regions indicate the ranges for which, by virtue of the addition of $\vartheta$, one can outperform the EF values at normal incidence. The EF values visited along a given $\mathcal{H}_i$ visits a periodic sequence of local peaks. We find that the distance, in $L$, between successive peaks is approximately equal to $\lambda_\text{SPP}/2$, resembling Fabry-Perot-like resonances with both even and odd parity modes being excited.

It must be noted that the complex patterns found here are beyond the simple Fabry-Perot picture. This becomes evident from a qualitative assessment of the plasmon modes excited across three distinct $\mathcal{H}_i$ families.  Figure~\ref{fig:Results4}(c) presents the electric field norm at the points labeled $M_1$, $M_2$, and $M_3$, all corresponding to $L = 1470$ nm. The $\mathcal{H}_1$ modes display a notable up-and-down differentiation, characterized by a low electric field magnitude along the lateral sides of the cylinder, while exhibiting a radially symmetric field distribution within the gap. In contrast, the $\mathcal{H}_2$ modes show a significant electric field magnitude along the lateral sides, with a more apparent triangular symmetry emerging within the gap region. These structural features remain consistent within the same family.

The appearance of the complex interference pattern, together with the nontrivial symmetry along the nanorod perimeter of the electric field modes at EF peaks, is consistent with the working regime. Namely, here the length $L$ and the diameter $D$ have values comparable to the wavelength. As mentioned above, the quasi-static approximation no longer suffices for analyzing plasmonic modes along the nanorod axis and the contribution of retardation effects becomes prominent. This enables the access to exciting multiple modes that may interfere. In this particular driving configuration there are two contributing synergic factors at place: (i) the presence of a wavevector component along the $z$-axis and (ii) the electric field polarization normal to the $z$-axis coexists with a wavevector component along such direction (here the $x$ direction). Both factors are absent in the polarization scenario of Sec.~\ref{sc:pola} where $\mathbf{k}\parallel \hat{x}$ and the field oscillation along a direction perpendicular to $z$ (in that case along $y$) is not able to excite modes producing hotspots at the gap. 

\begin{figure}[!t]
    \centering    \includegraphics[width=0.9\textwidth]{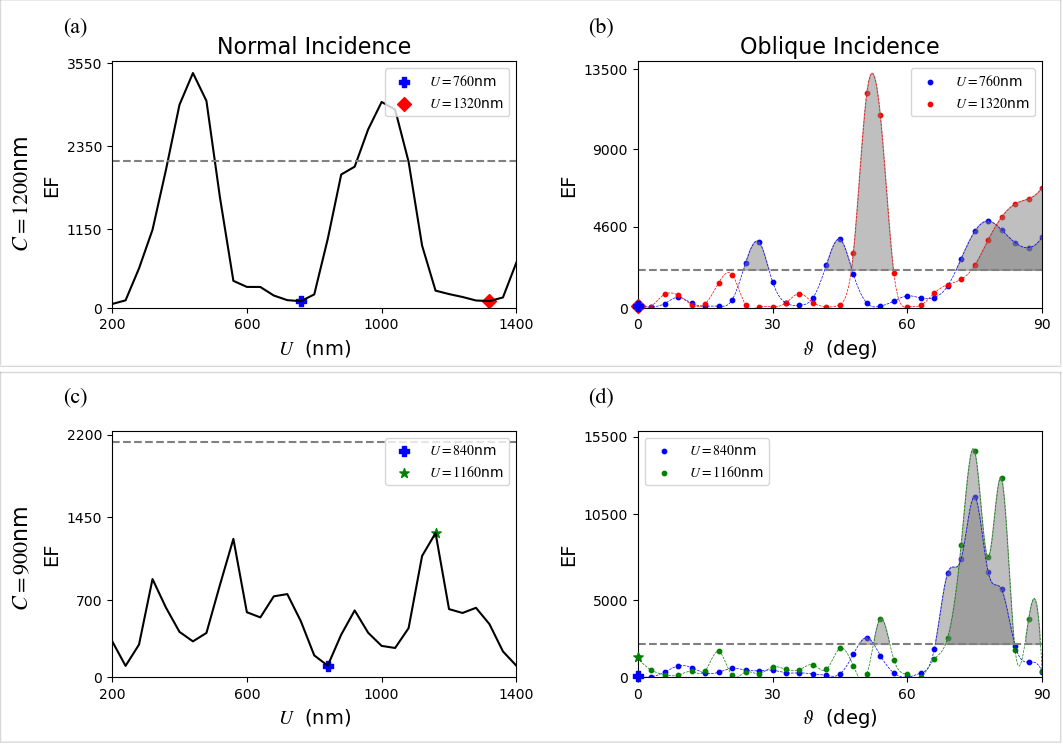}
    \caption{Enhancement Factor at the gap versus $U$ for a nanorod dimer with diameter $D=360$ nm for covered length (a) $C=1200$ nm or (b) $C=900$ nm under normal incidence. These EF can be greatly reduced if $U$ is not properly tuned. Symbols mark specific values of $U$ that obtain low EF at normal incidence; at these conditions panels (b) and (d) present the enhancement factor dependence with the angle of incidence $\vartheta$. The dashed line represents the maximum baseline EF achieved under normal incidence for the uncovered case. The shaded areas over such baseline level highlight that significant EF values can be achieved by adjusting $\vartheta$ even for the most unfavorable cases.}
    \label{fig:Results5}
\end{figure}

Building on the above described success of this strategy for mitigating EF minima in uncovered configurations, we now briefly inspect the covered case (Fig.~\ref{fig:Results2}(a)) to assess the effectiveness of varying the inclination angle $\vartheta$. Specifically, we focus on the two exemplary covered length cases shown in Fig.~\ref{fig:Results2}, which previously exhibited high ($C=1200$ nm) and low ($C=900$ nm) EF maxima for the $D=360$ nm case when $\vartheta=0$. In Fig.~\ref{fig:Results5}(a) and (c), we show the computed EF profile versus the uncovered length $U$. For both cases there are ranges of $U$ that produce low EF, being inefficient for SERS operation at normal incidence. On the other hand, in cases that $U$ is well tuned, large EF at normal incidence are generated and there is no need to proceed to increment $\vartheta$. However, we have verified that in those cases this strategy can also increase the EF even higher (not shown).

Here, instead, we show the behavior of this approach for at values of $U$ (see symbols at Fig.~\ref{fig:Results5}(a) and (c)) that produce the worst EF at normal incidence. The results, shown in Fig.~\ref{fig:Results5}(b) and (d), demonstrate that, for each of these configurations, there exists several values of the angle $\vartheta$ at which the computed EF surpasses the maximum value (the horizontal dashed line at all panels of Fig.~\ref{fig:Results5}) found when the uncovered case at normal incidence is tuned to an optimal length $L$. The ranges of $\vartheta$ that fulfill this goal are plotted as shaded regions in Fig.~\ref{fig:Results5}(b) and (d). In addition, the obtained EF can also outperform the increased EF value obtained for the optimal $U$ at $\vartheta=0$ in a covered sample, i.e., the peak values in Fig.~\ref{fig:Results5}(a). Therefore, these results confirm that adjusting the angle of incidence is a viable strategy to mitigate EF minima also in the covered design of Fig.~\ref{fig:Results2}(a).

\section{Conclusions}\label{SC:Conclusions}

Previous studies showed that the SERS enhancement factor at the gap of gold nanorod dimers, when illuminated by a He-Ne CW laser of $\lambda_0=633$ nm, can be greatly suppressed if the lengths of the two nanorods are not carefully chosen. Thus, either fabrication errors or constructing the device at arbitrary lengths can cause it to malfunction as an efficient SERS detector operating at such widely available laser frequency. Here, we have presented two strategies to overcome this limitation. These strategies leverage the retardation regime, enabling the utilization of other SPP modes that also produce significant enhancement of the near field at the gap.

The first successful strategy involves maintaining the customary normal incidence excitation configuration ($\mathbf{k}\parallel \hat{x}$ and $\mathbf{E}_0\parallel \hat{z}$, corresponding to Fig.~\ref{fig:setup}(e) with $\varphi=0$) while reducing the diameter to $260$ nm. We show that, within the entire working range of the length $L$, different modes are excited that not only avoid the EF minima but also surpass the peak values obtained for the $D=360$ nm case. We then presented a second strategy that achieves large EF values across the relevant $L$ range while retaining the $D=360$ nm diameter from previous studies. This is accomplished at the measurement stage by increasing the incidence angle $\vartheta$, which introduces a wavevector component along the nanorods' $z$-axis while keeping the electric field polarization within the $xz$-plane (i.e., TM polarized, as sketched in Fig.~\ref{fig:setup}(f)). We show that as the phase of the incident wave becomes $z$-dependent, given that $L$ is comparable to $\lambda_0$, retardation effects allow for $\vartheta$-selective excitation of modes hosting large EF values at the gap.

In summary, these results open the possibility of obtaining SERS measurements of similar or even greater quality than those reported experimentally in previous studies~\cite{pedano,shuzho}, while removing the constraint of having to fabricate the device at precise values of the nanorods' length $L$. Importantly, our simulations show that both strategies apply successfully to the covered design, in which the far-from-the-gap ends of the nanorods are covered by a layer of gold~\cite{IvanRSC2021}. Since the covering procedure introduces fabrication errors, the fact that the SERS EF can be enhanced regardless of the uncovered length $U$ is a significant advantage for the prospect of devising a hybrid system with complementary electric transport measurements.

\backmatter

\bmhead{Acknowledgements}

We thank M.L. Pedano for useful discussions. 

\section*{Declarations}

\bmhead{Funding}
This work has been supported by the Consejo Nacional de Ciencia, Tecnología e Innovación Tecnológica del Perú (CONCYTEC), Contract N° 174-2018-FONDECYT-BM.  A.A.R acknowledges support by PAIDI 2020 Project No. P20-00548 with FEDER funds, additional support from CONICET (Argentina) and funding from FONCyT, PICT-2016-2531, PICT-2020-SERIEA-02705 (Argentina).

\newpage
\begin{appendices}

\section{EF$(L)$ for a range of nanorod diameters}\label{SC:Appendix1}
\begin{figure}[!b]
    \centering  \includegraphics[width=0.9\textwidth]{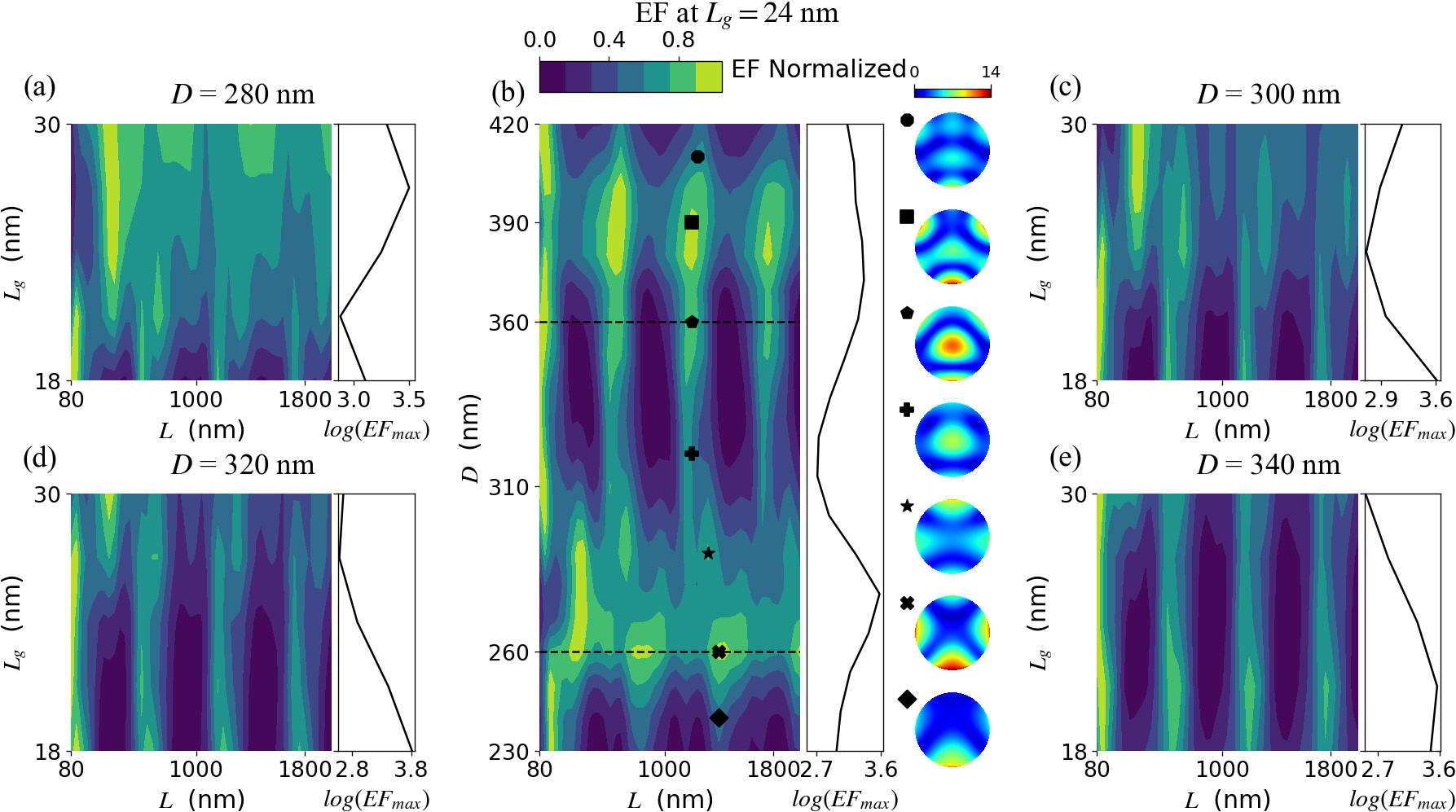}
    \caption{Colormaps of enhancement factor as a function $L$ and $L_g$ for (a) $D=280$ nm, (c) $D=300$ nm, (d) $D=320$ nm and (e) $D=340$ nm where field intensities are normalized across each gap size. Panel (b) offers an overview of the EF profile for a range of $D$ values between $230$ nm and $390$ nm. Additionally, electric field norm profiles are shown as insets for selected values of $D$ following a local maximum. Dashed lines indicate the cases presented in Fig.~\ref{fig:Results1}.}
    \label{fig:Appendix1}
\end{figure}
In Sec.~\ref{sc:diameter}, we highlighted the case of $D=260$ nm as an exemplary configuration for achieving consistently high EF across a broad range of $L$. To support this claim, panel (b) in Fig.~\ref{fig:Appendix1} shows the EF profile as a function of diameter at a fixed gap of $L_g=24$ nm. When normalized, it becomes apparent that the $D=260$ nm configuration distinguishes itself by exhibiting a EF profile free from low values of EF as observed in other diameter settings, while still not diminishing the periodic local maxima, as opposed to nearby diameters such as $D=270$ nm or $D=250$ nm. 

Further comparisons are provided in Fig.~\ref{fig:Appendix1} panels (a), (c), (d), and (e), where the results for alternative diameter and gap combinations are displayed. These additional cases reinforce our earlier conclusion, showcasing that while other configurations exhibit significant oscillations in EF with variations in $L$, the $D=260$ nm case consistently mitigates such minima, accentuating its suitability for stable SERS performance across a range of practical antenna dimensions. In addition, from Fig.~\ref{fig:Appendix1}(b) we note that around $D=390$ nm the EF profile also develops less pronounced minima. Despite the fact that in that case the EF values are a factor of 2 smaller than in the $D=260$ nm case, dimers with $D=390$ nm could also enable SERS operation not requiring fine tuning of the nanorod length $L$.

Fig.~\ref{fig:Appendix1}{(b)} also includes as insets the electric field norm profile at the gap for different diameters along a EF local maximum. This reveals the shape at the gap of the dominant multipolar mode being excited~\cite{Lidong} . 
Specifically, for $D=360$ nm, the excited mode develops a dominant lobe at the center of the circular section whereas for $D=260$ nm the field norm exhibits a quadrupolar distribution with four distinct regions of high amplitude along the edges, separated by diagonal zones of low amplitude. This indicates a change in the modal structure as the diameter increases from 260 nm to 360 nm, correlating with a change in the EF profile with respect to the nanorods' length.

\section{On slight left/right cylinder asymmetries in the EF for $\vartheta\neq 0$} \label{SC:Appendix2}
\begin{figure}[!b]
    \centering \includegraphics[width=0.9\textwidth]{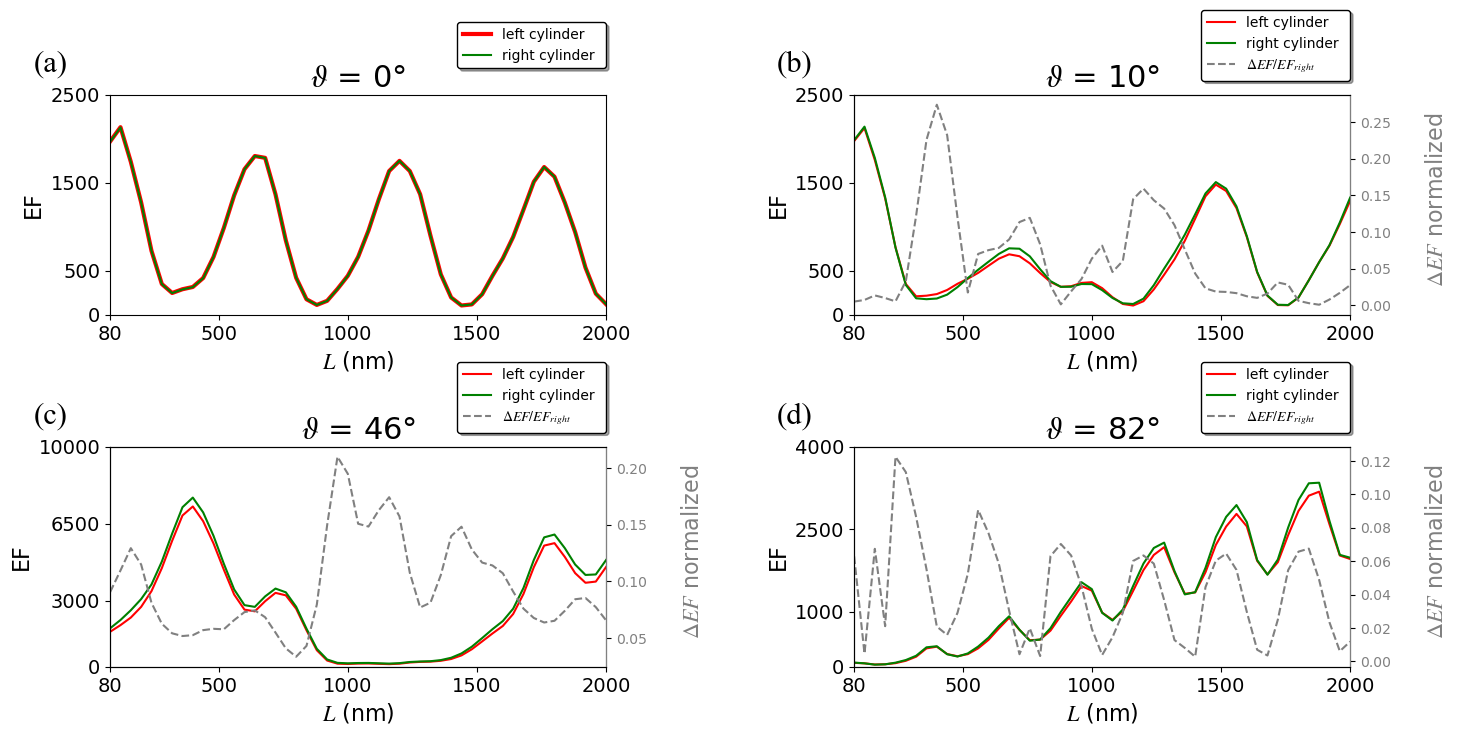}
    \caption{Enhancement Factor asymmetries as a function of nanorod length $L$ for (a)$\vartheta=0$, (b)$\vartheta=10^\circ$, (c)$\vartheta=46^\circ$ and (d)$\vartheta=82^\circ$, shown for both the left (red line) and right (green line) cylinders. The normalized difference (gray dashed line) quantifies the asymmetry in EF between the cylinders as the incidence angle increases. At normal incidence, both profiles overlap.}
    \label{fig:Appendix3}
\end{figure}
For non-zero incidence angles ($\vartheta\neq 0$), the obtained EF at the gap are different if the computation of Eq.~\ref{eq:EF} is taken over a surface $2$ nm away from the face of the left ($z < 0$) or right ($z > 0$) cylinder. Figure~\ref{fig:Appendix3} depicts how varying the incidence angle $\vartheta$ disrupts the symmetry and produce different EF values. Despite this variation in magnitude, the asymmetry, which in the worst case is $20$\%, does not alter the fundamental excitation of SPP modes, as both cylinders retain the same overall EF profile shape. At normal incidence ($\vartheta=0$), this difference is negligible, with relative error falling below 0.1\% due to numerical precision limitations, and thus is not plotted. For practical purposes, in Sec.~\ref{sc:strategy3}, we present the minimum of the two values, providing a conservative estimate of the effectiveness of the proposed method.

\begin{figure}[t]
    \centering \includegraphics[width=0.9\textwidth]{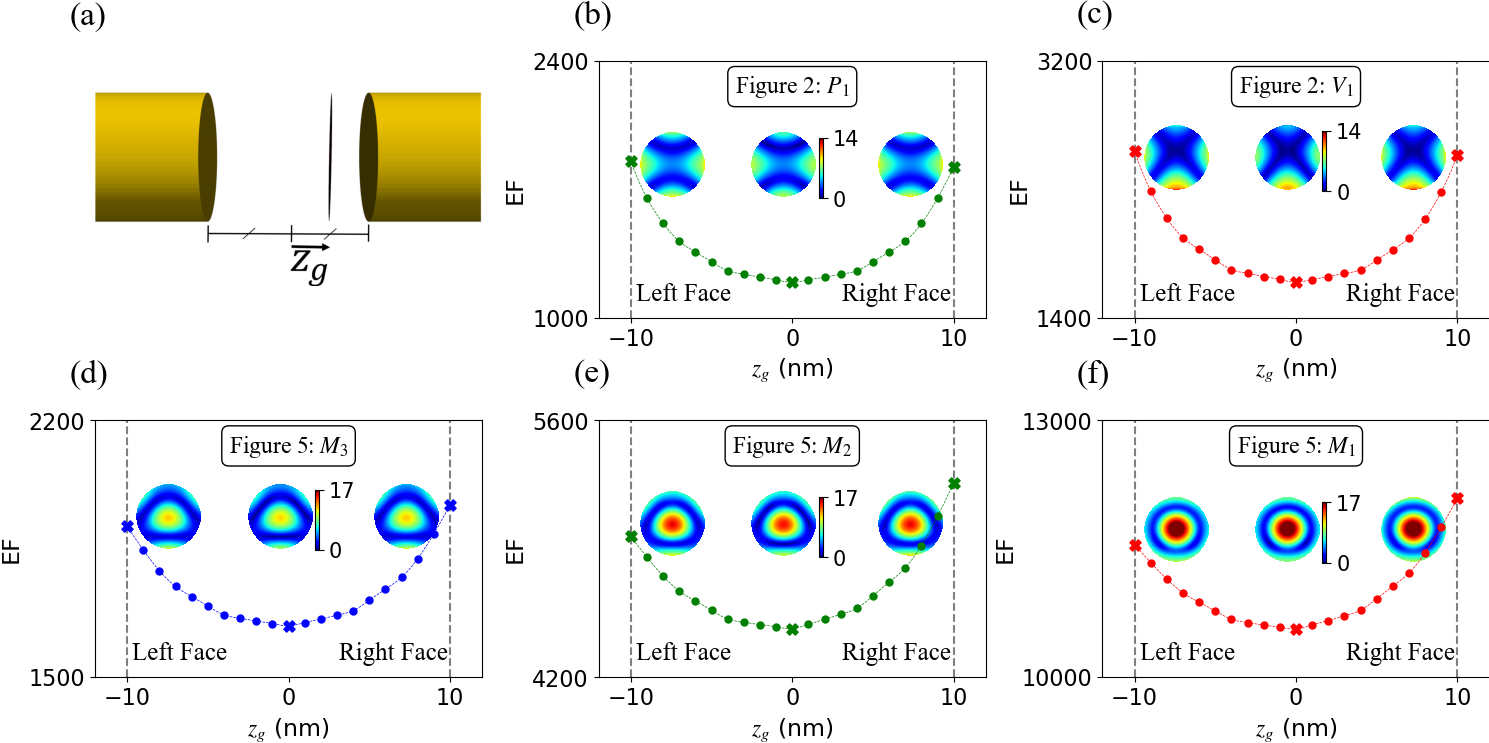}
    \caption{Enhancement factor calculated at different $z_g$: the distance from the center of the gap defined in (a). Panels (b) and (c) depict the EF variation with $z_g$ for $D=260$ nm at peaks $P_1(L=1200\text{ nm})$ and $V_1(L=1440\text{ nm})$, respectively. Panels (d), (e), and (f) illustrate the left-right asymmetries in the EF for the points $M_3$, $M_2$ and $M_1$ respectively, defined in Fig.~\ref{fig:Results4} as the intersection of the hyperbolas $\mathcal{H}_3$, $\mathcal{H}_2$, $\mathcal{H}_1$ with $L=1470$ nm. Insets in each panel display the electric field norm profiles for $z_g=-10, 0, 10$ nm, corresponding to points marked by "x" symbols in the scatter plots.}
    \label{fig:Appendix4}
\end{figure}

Figure~\ref{fig:Appendix4} shows the EF calculated at other positions inside the gap $z_g$, which is measured from the center of the gap. Note that $z_g=-10$ nm and $z_g=10$ nm correspond to the situations of Fig.~\ref{fig:Appendix3} being the EF computed $2$ nm away from the left and right nanorod surface, respectively. We only include exemplary points to depict the generic behavior related to the two successful strategies discussed in the main-text. The EF dependence within the gap for the $D=260$ nm case at normal incidence is included in panels (b) and (c). In these cases, the enhancement factor has even symmetry about $z_g=0$ with its magnitude being slightly weaker at the center of the gap. Exemplary points for the $D=360$ nm oblique incidence case are included in panels (d), (e), and (f) corresponding to $\vartheta=12^\circ$, $37^\circ$, and $74^\circ$, respectively. As the angle of incidence increases, the even symmetry breaks, and the enhancement factor becomes stronger on the right side of the gap compared to the left. Interestingly, the rate at which the EF diminishes with distance from the cylinder faces decreases as the tilt angle of the incident electric field grows. Despite these weak asymmetries, the electric field profile is virtually the same throughout the gap.

\section{Parameters for the $\mathcal{H}$ and $\mathcal{V}$ families}\label{SC:Appendix3}

Each family aligns with a set of parallel hyperbolic trends, represented as follows:
\begin{align}
    \text{$\mathcal{V}_i:\,\,\,\,\,$    } \vartheta^{\mathcal{V}}_i\!\!(L)  &=   \frac{V}{L - \ell^{\mathcal{V}}_i}-\theta^{\mathcal{V}}_i \,\,\,\,\,\,(\text{deg}). \label{eq:phiV}\\
    \text{$\mathcal{H}_i:\,\,\,\,\,$    } \vartheta^{\mathcal{H}}_i\!\!(L) &= \theta^{\mathcal{H}}_i - \frac{H}{L + \ell^{\mathcal{H}}_i}  \,\,\,\,\,\,(\text{deg}), \label{eq:phiH}
\end{align}
 Table \ref{tab:HiperbolaDefinition} lists the parameters for these lines as presented in Sec.~\ref{sc:strategy3}. The parameters  $\ell^\mathcal{X}_i$'s are given in nanometers, the  $\theta^{\mathcal{X}}_i$ in deg, while both $H$ and $V$ are given in $\text{deg}\cdot \text{nm}$.

\begin{table}[t]
    \caption{Parameters of hyperbolas $\mathcal{H}_i$ and $\mathcal{V}_i$ defined in Eqs.~\ref{eq:phiV} and \eqref{eq:phiH} and plotted in Fig.~\ref{fig:Results4}.}
    \centering
        \begin{tabular}{|c|c|c|}
        \cmidrule{1-3}
        $\mathcal{V}_1$ & $\ell^{\mathcal{V}}_1 = -99.0$, $\theta^{\mathcal{V}}_1 = 83.3$ & \\ \cmidrule{1-2}
        $\mathcal{V}_2$ & $\ell^{\mathcal{V}}_2 = 130$, $\theta^{\mathcal{V}}_2 = 58.3$ & $V = 63.0 \times 10^3$\\ \cmidrule{1-2}
        $\mathcal{V}_3$ & $\ell^{\mathcal{V}}_3 = 386$, $\theta^{\mathcal{V}}_3 = 45.4$ & \\ \cmidrule{1-3}
        $\mathcal{H}_1$ & $\ell^{\mathcal{H}}_1 = 895$, $\theta^{\mathcal{H}}_1 = 107$ &\\ \cmidrule{1-2}
        $\mathcal{H}_2$ & $\ell^{\mathcal{H}}_2 = 394$, $\theta^{\mathcal{H}}_2 = 79.2$  & $H = 78.2\times 10^3$\\ \cmidrule{1-2}
        $\mathcal{H}_3$ & $\ell^{\mathcal{H}}_3 = 79.0$, $\theta^{\mathcal{H}}_3 = 63.0$ & \\ \cmidrule{1-3}
    \end{tabular}
    \label{tab:HiperbolaDefinition}
\end{table}

\end{appendices}

\bibliographystyle{sn-mathphys-num}
\bibliography{sn-bibliography}

\end{document}